\documentclass{iaus}
\usepackage{graphicx}

\def\aa{\textit{A\&A}\ }

\def\apj{\textit{ApJ}\ }
\def\apjs{\textit{ApJS}\ }
\def\mnras{\textit{MNRAS}\ }
\def\nat{\textit{Nature}\ }
\def\lsim{\mathrel{\rlap{\lower 4pt \hbox{\hskip 1pt $\sim$}}\raise 1pt
\hbox {$<$}}} 
\def\gsim{\mathrel{\rlap{\lower 4pt \hbox{\hskip 1pt $\sim$}}\raise 1pt
\hbox {$>$}}}
\newcommand{\Msun}{M_\odot}

\newcommand{\Nifs}{$^{56}$Ni}
\newcommand{\Mni}{M{\rm (^{56}Ni)}}
\newcommand{\Ed}{\dot{E}_{\rm dep}}
\newcommand{\Edep}{\dot{E}_{\rm dep,51}}

\title[Supernova Yields] 
{Chemical Yields from Supernovae and Hypernovae}

\author[Ken'ichi Nomoto {\textit et al.}]   
{Ken'ichi~Nomoto$^{1,2}$, Shinya~Wanajo$^{1,2}$, Yasuomi~Kamiya$^{1,2}$,
Nozomu~Tominaga$^3$, \and Hideyuki~Umeda$^2$}

\affiliation{$^1$Institute for the Physics and Mathematics of the
Universe, University of Tokyo\\Kashiwanoha 5-1-5, Kashiwa, Chiba
277-8568, Japan\\ email: {\tt nomoto@astron.s.u-tokyo.ac.jp}
\\[\affilskip]
$^2$Department of Astronomy, University of Tokyo, Bunkyo-ku, Tokyo
113-0033, Japan \\
$^3$National Astronomical Observatory, Mitaka, Tokyo 113-0033, Japan}

\pubyear{2008}
\volume{254}  
\pagerange{16}
\jname{The Galaxy Disk in Cosmological Context}
\editors{J. Andersen, J. Bland-Hawthorn \& B. Nordstr\"{o}m, eds.}

\begin{document}

\maketitle

\begin{abstract}
We review the final stages of stellar evolution, supernova
properties, and chemical yields as a function of the progenitor's mass
$M$.
(1) 8 - 10 $M_\odot$ stars are super-AGB stars when the
O+Ne+Mg core collapses due to electron capture.  These AGB-supernovae
may constitute an SN 2008S-like sub-class of Type IIn supernovae.
These stars produce little $\alpha$-elements and Fe-peak elements, but
are important sources of Zn and light p-nuclei.
(2) 10 - 90 $M_\odot$ stars undergo Fe-core collapse.  Nucleosynthesis
in aspherical explosions is important, as it can well reproduce the
abundance patterns observed in extremely metal-poor stars.
(3) 90 - 140 $M_\odot$ stars undergo pulsational nuclear instabilities 
at various nuclear burning stages, including O and Si-burning.
(4) 140 - 300 $M_\odot$ stars become pair-instability supernovae, if
the mass loss is small enough.
(5) Very massive stars with $M \gsim 300 M_\odot$ undergo
core-collapse to form intermediate mass black holes.

\keywords{Galaxy: halo
--- gamma rays: bursts 
--- nuclear reactions, nucleosynthesis, abundances 
--- stars: abundances --- stars: AGB
--- supernovae: general}
\end{abstract}

\firstsection 
\section{Core-Collapse Supernovae and Progenitor Masses}

The final stages of massive star evolution, supernova properties, and
their chemical yields depend on the progenitor's masses $M$ as follows
(e.g., \cite{arnett1996}):

(1) {\sl 8 - 10 $M_\odot$ stars}: These stars are on the AGB phase
when the O+Ne+Mg core collapses due to electron captures.  The exact
mass range depends on the mass loss during the AGB phase.  They
undergo weak explosions being induced by neutrino heating.  These
stars produce little $\alpha$-elements and Fe-peak elements, but are
important sources of Zn and light p-nuclei.  These AGB supernovae may
constitute an SN 2008S-like sub-class of Type IIn supernovae.

(2) {\sl 10 - 90 $M_\odot$ stars}: These stars undergo Fe-core
collapse to form either a neutron star (NS) or a black hole (BH), and
produce a large amount of heavy elements from $\alpha$-elements and
Fe-peak elements.  Observations have shown that the explosions of
these Fe-core collapse supernovae are quite aspherical.  In the
extreme case, the supernova energy is higher than $10^{52}$ erg
s$^{-1}$, i.e. a Hypernova.  Nucleosynthesis in these jet-induced
explosions is in good agreement with the abundance patterns observed
in extremely metal-poor stars.

(3) {\sl 90 - 140 $M_\odot$ stars}: These massive stars undergo
nuclear instabilities and associated pulsations ($\epsilon$-mechanism)
at various nuclear burning stages depending on the mass loss and thus
metallicity.  In particular, if the mass loss is negligible,
pulsations of O-cores and/or Si-cores due to O, Si-burning could
produce dynamical mass ejection.  Eventually, these stars undergo
Fe-core collapse.

(4) {\sl 140 - 300 $M_\odot$ stars}: If these very massive stars (VMS)
do not lose much mass, they become pair-instability supernovae (PISN).
The star is completely disrupted without forming a BH and thus ejects a
large amount of heavy elements, especially Fe.

(5) {\sl Stars with $M \gsim 300 M_\odot$}: These VMSs are too massive
to be disrupted by PISN but undergo core collapse (CVMS), forming an
intermediate-mass black hole (IMBH).  Some mass ejection could be
possible, associated with the possible jet-induced explosion.

Here we summarize the properties of the above supernovae and chemical
yields in some detail.

\begin{figure}[t]
\begin{center}
\includegraphics[width=9.5cm]{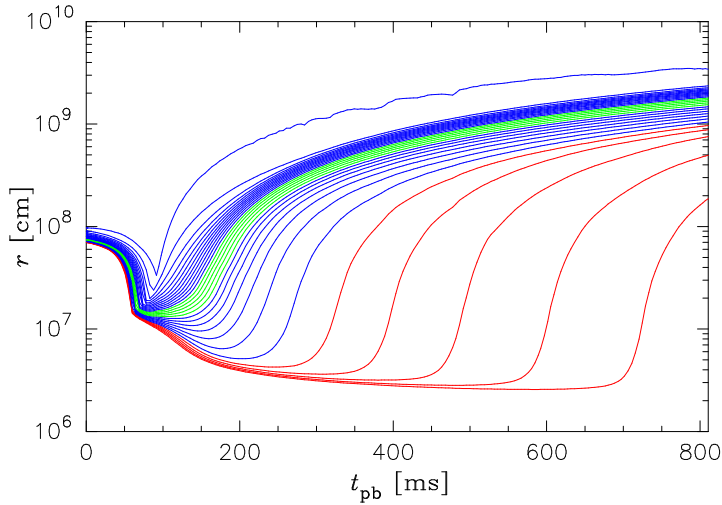}
\end{center}
\caption{
The change in the radius as a function of post-bounce time for
material ejected from the collapsing O-Ne-Mg core
(\cite{kitaura06}).
}
\label{fig:9mrt}
\end{figure}

\section{8 - 10 $M_\odot$ AGB Stars undergoing Electron Capture Supernovae}

\subsection{Collapse of O+Ne+Mg Core induced by Electron Capture}

     For 8 - 10 $M_\odot$ stars, electrons become degenerate already
in a C+O core.  In a semi-degenerate C+O core, neutrino cooling leads
to off-center ignition of carbon when the C+O core mass exceeds the
critical mass of 1.06 $M_\odot$.  The off-center carbon burning shell
moves inward all the way to the center due to heat conduction
(\cite{nom84}; \cite{timmes92}; \cite{berro97}).

     After exhaustion of carbon in the central region, an O+Ne+Mg core
forms.  The core mass does not exceed the critical mass of 1.37
$M_\odot$ for neon ignition and, hence, neon burning is never ignited
(\cite{nom84}).  Then the O+Ne+Mg core becomes strongly degenerate.
The envelope becomes similar to the asymptotic giant-branch
(super-AGB) stars (\cite{hashi93}; \cite{poel08}) with a thin He
burning shell that undergoes thermal pulses and $s$-process
nucleosynthesis.

     The final fate depends on the competition between the mass loss
that reduces the envelope mass and the increase in the core mass
through the H-He shell burning.  If the mass loss is fast, an O+Ne+Mg
white dwarf is formed, which could be the case for 8 $M_\odot$ -
M$_{\rm up}$ stars, where M$_{\rm up} \sim 9 \pm 0.5 M_\odot$ being
smaller for smaller metallicity (\cite{poel08}).  For M$_{\rm up}$ -
10 $M_\odot$ stars, the core mass grows to 1.38 $M_\odot$ and the
central density reaches 4 $\times$ 10$^9$ g cm$^{-3}$.  The electron
Fermi energy exceeds the threshold for electron captures
$^{24}$Mg(e$^-,\nu$) $^{24}$Na (e$^-,\nu$) $^{20}$Ne and $^{20}$Ne
(e$^-,\nu$) $^{20}$F (e$^-,\nu$) $^{20}$O.  The resultant decrease in
$Y_e$ triggers collapse (\cite{nom87}).

     The hydrodynamical behavior of collapse and bounce is somewhat
different from the iron core collapse of more massive stars
(Fig.~\ref{fig:9mrt}: \cite{kitaura06}).  The explosion
energy is as low as $E \sim 10^{50}$ erg.  The explosion is also
suggested be close to spherical, thus producing little pulsar kick.
The existence of pulsars in globular clusters might be explained by
the electron capture supernovae (\cite{kalgera08}).

\begin{figure}[t]
\begin{center}
\includegraphics[width=9.0cm]{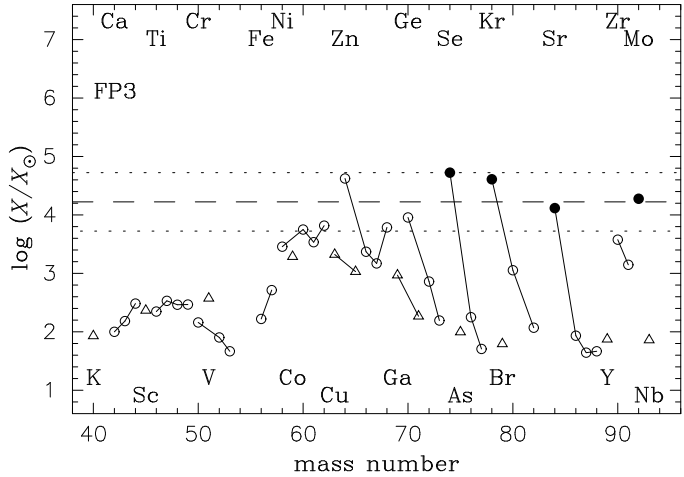}
\end{center}
\caption{
Mass fractions of isotopes (after decay) in the ejecta of model FP3
(\cite{wana08}) relative to their solar values (\cite{Lodd03}) as a
function of the mass number. The even-$Z$ and odd-$Z$ isotopes are
denoted by open circles and triangles, respectively. The $p$-nuclei
are represented by filled symbols. The dotted horizontal lines
indicate a ``normalization band'' between the largest production
factor and a factor of ten smaller than that, along with the median
value (\sl{dashed line}).
}
\label{fig:9mabund}
\end{figure}

\subsection{Nucleosynthesis in Electron Capture Supernovae}

Nucleosynthesis in the supernova explosion of a $9 \, M_\odot$ star
has been investigated (\cite{hofman08}; \cite{wana08}) using
thermodynamic trajectories taken from the explosion model
(\cite{kitaura06}).  Here we summarize the results by Wanajo \etal\
(2009).

1. The unmodified model produces small amounts of
$\alpha$-elements and iron, but large amounts of $^{64}$Zn,
$^{70}$Ge, and in particular, $^{90}$Zr, with some light $p$-nuclei
(e.g., $^{92}$Mo; Fig.~\ref{fig:9mabund}).  This is due to the
ejection of a large amount of neutron-rich matter ($Y_e =
0.46-0.49$), and might put severe constraints on the frequency of 
occurrence of this type of supernovae (\cite{hofman08}).  However, the
production of $^{90}$Zr does not serve as a strong constraint, because
it is easily affected by a small variation of $Y_e$ (see below).

2. The overproduction of $^{90}$Zr becomes more moderate if the minimum $Y_e$ is
only $1-2\%$ larger than that in the unmodified model.  Such a change
in the initial $Y_e$ profile might be caused by convection that is not
considered in the 1-D simulation (\cite{kitaura06}).  In this case
(model FP3: Fig.~\ref{fig:9mabund}), the largest overproduction, which
is shared by $^{64}$Zn, $^{70}$Se, and $^{78}$Kr, falls to one-tenth
that of the unmodified model. The $^{64}$Zn production provides an
upper limit to the occurrence of exploding O-Ne-Mg cores at 
about $20\%$ of all core-collapse supernovae.

3. The ejecta mass of $^{56}$Ni is $0.002-0.004\, M_\odot$, much
smaller than the $\sim 0.1\, M_\odot$ in more massive progenitors.  
Convective motions near the mass cut may also affect the
$Y_e$-distribution and thus the $^{56}$Ni mass.  See Wanajo \etal\ (2008)
for a recent comparison between the electron capture supernova yields
and abundances in the Crab Nebula.

\begin{figure}[t]
\begin{center}
\includegraphics[width=10.0cm]{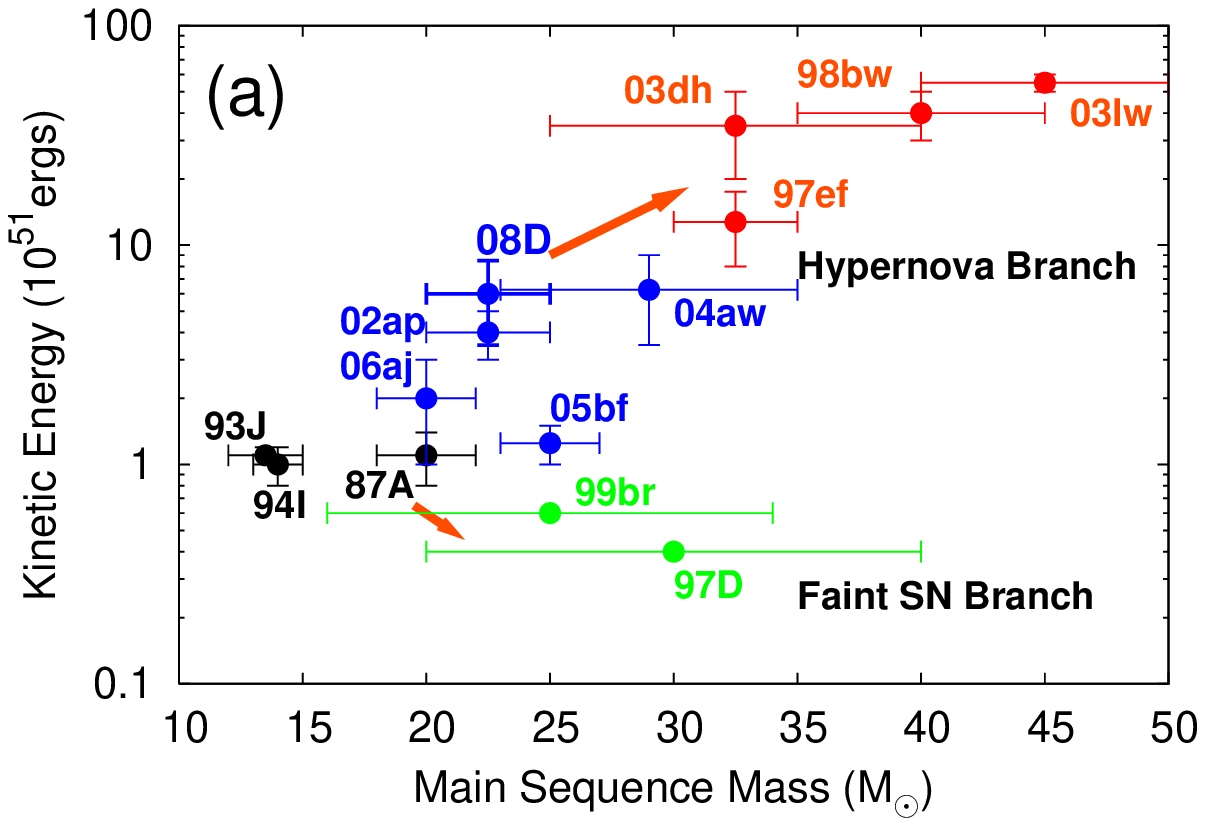}
\includegraphics[width=10.0cm]{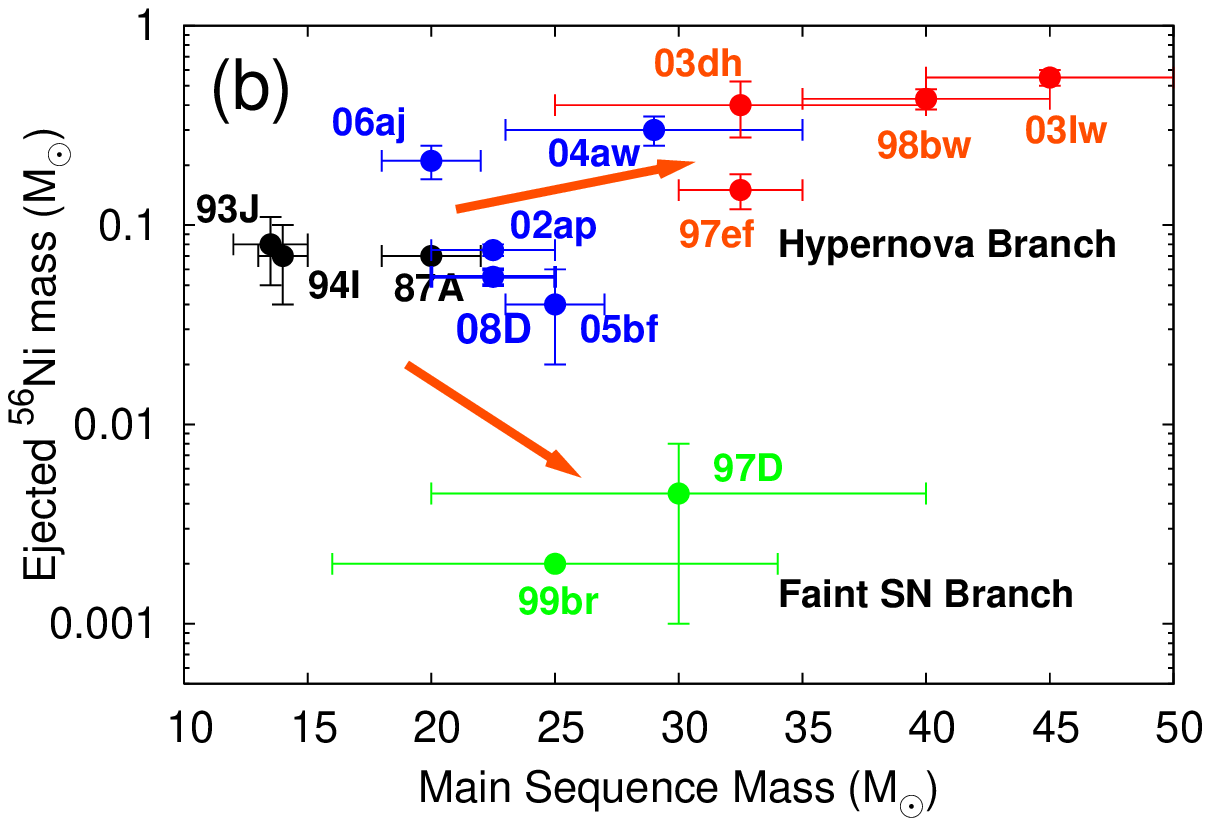}
\end{center}
\caption{
{\it Upper}: The kinetic explosion energy $E$ as a function of the main
sequence mass $M$ of the progenitors for several
supernovae/hypernovae. Hypernovae are the SNe with $E_{51}>10$.
{\it Lower}: Same as the upper panel, but for the ejected mass of
$^{56}$Ni.
}
\label{figME}
\end{figure}

\subsection{Connection to Faint Supernovae}

The expected small amount of $^{56}$Ni as well as the low explosion
energy of electron capture supernovae have been proposed as an
explanation of the observed properties of low-luminosity SNe~IIP, such
as SN~1997D (\cite{Chug00}; \cite{kitaura06}) and of the low
luminosity of SN~2008S-like transients (\cite{prieto08};
\cite{thom08}). The estimated $^{56}$Ni masses of $\sim 0.002-0.008\,
M_\odot$ for observed low-luminosity SNe~II-P (\cite{Zamp03};
\cite{Hend05}) are in reasonable agreement with the presented results
from O-Ne-Mg core explosions. An alternative possibility of such
supernovae is that more massive stars ($\gsim 20\, M_\odot$) with low
explosion energies suffer from fallback of freshly synthesized
$^{56}$Ni (\cite{Tura98}; \cite{nom03}; \cite{Zamp03}). A recent
analysis of the progenitors of SNe~IIP by \cite{Smar08} favors low
mass progenitors.  A lack of $\alpha$-elements such as O and Mg in the
case of collapsing O-Ne-Mg cores will be a key to spectroscopically
distinguish between these two scenarios.

Recently, the progenitor of SN 2008S was discovered in the infrared,
and it has been suggested that it was an AGB star (\cite{prieto08}).  
If so, SN 2008S could belong to the electron capture supernovae.

The envelope of the AGB star is carbon-enhanced (\cite{nom87}). Then
dust could easily be formed to induce mass loss. This may result in a
deeply dust-enshrouded object such as the progenitor of SN~2008S
(\cite{prieto08}; \cite{thom08}).  This might also imply that the mass
range of the stars that end their lives as O-Ne-Mg supernovae is 
$\sim 9.5-10\, M_\odot$; their frequency, $\sim 7-8\%$ of all the
core-collapse events, satisfies the constraint from our
nucleosynthesis results ($< 30\%$).

\begin{figure}[t]
\begin{center}
\includegraphics[width=6.7cm]{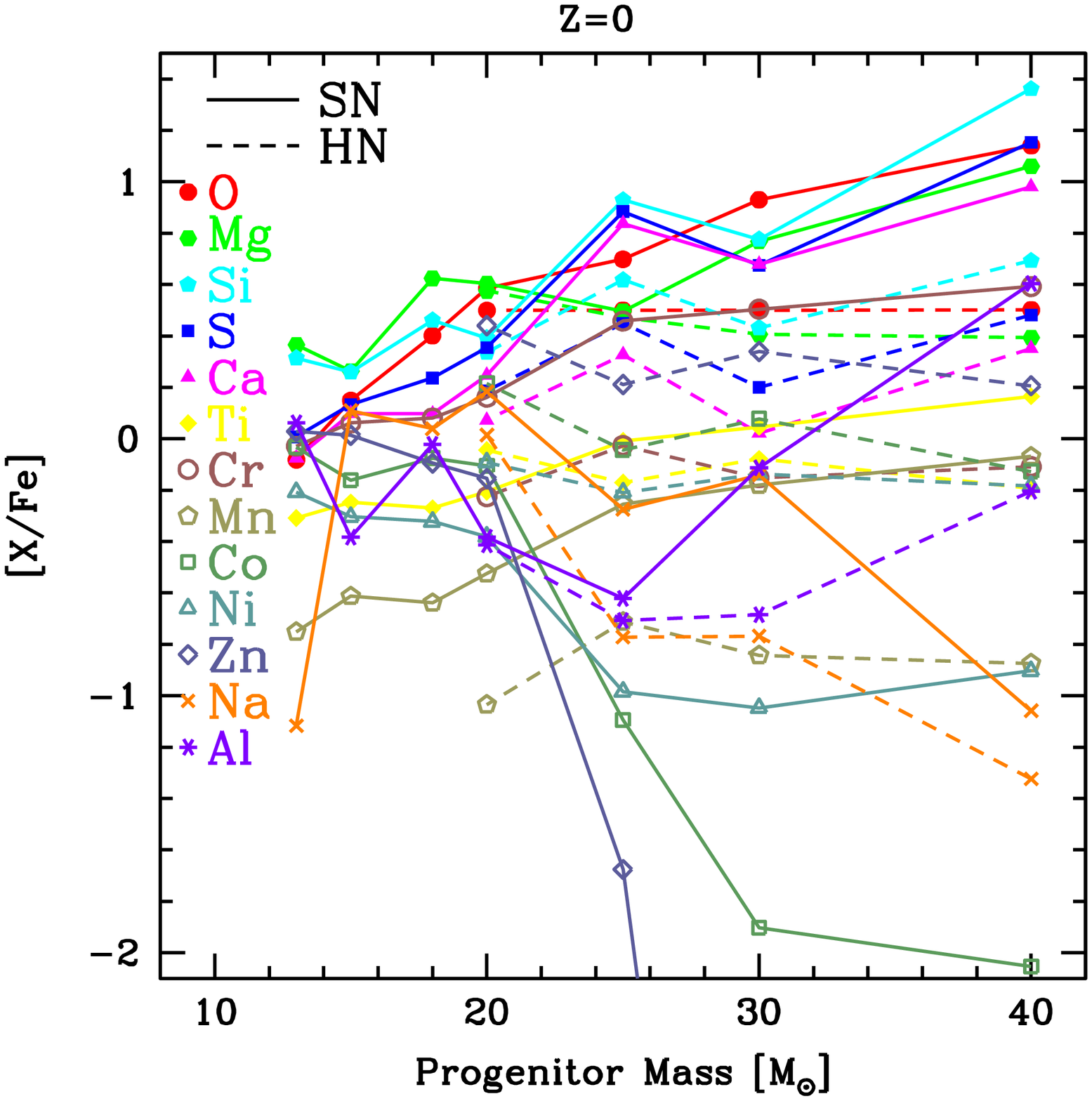}
\includegraphics[width=6.7cm]{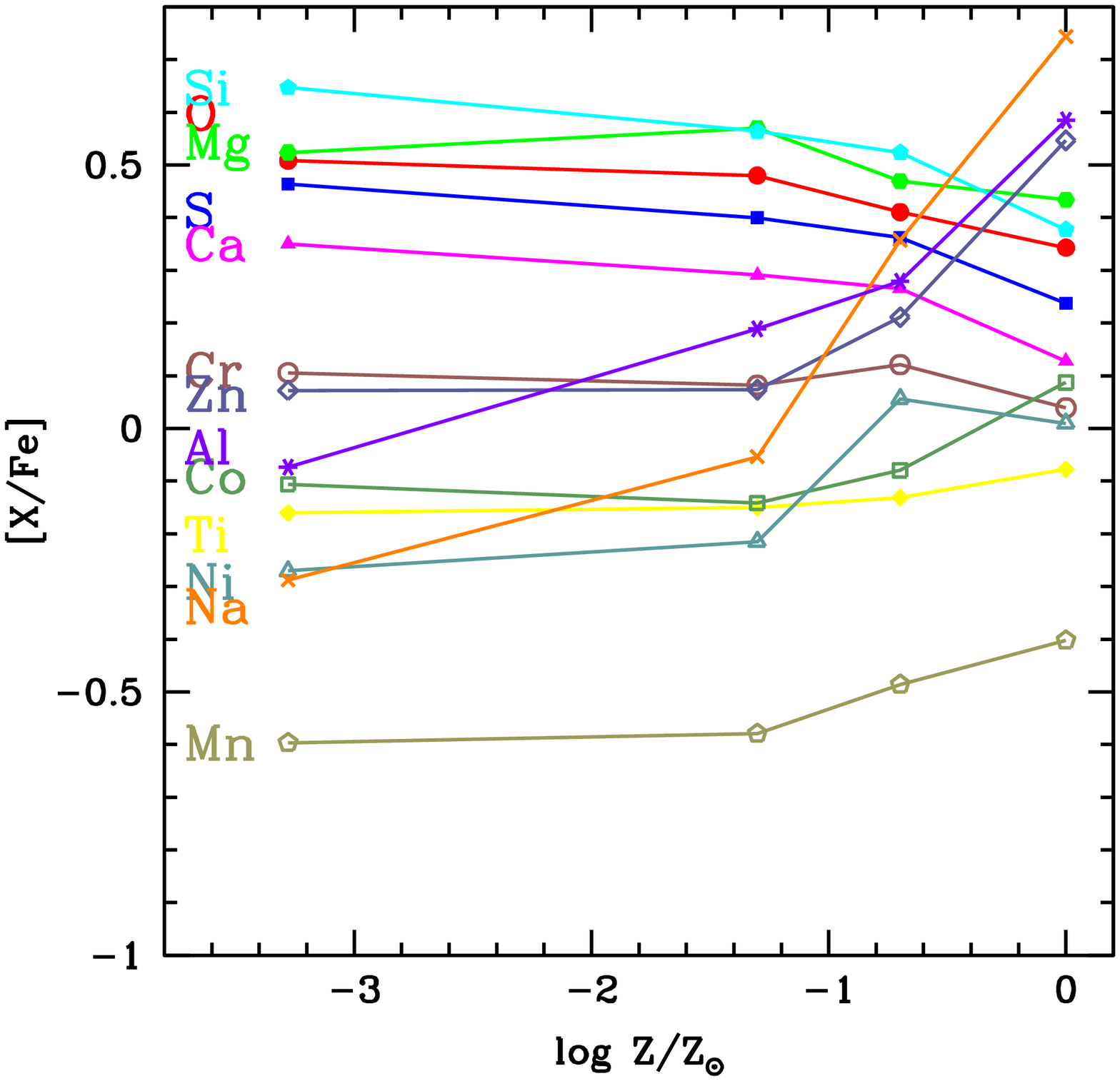}
\end{center}
\caption{({\sl Left}:) Relative abundance ratios as a function of progenitor 
mass with $Z=0$.  The solid and dashed lines show normal SNe II with
$E_{51}=1$ and HNe.  ({\sl Right}:) The IMF weighted abundance ratios as a
function of metallicity of progenitors, where the HN fraction
$\epsilon_{\rm HN}=0.5$ is adopted.  Results for $Z=0$ are plotted at
$\log Z/Z_\odot=-4$ (\cite{nomoto2006}; \cite{koba06}).}
\label{fig:yield}
\end{figure}

\section{10 - 90 $M_\odot$ Stars undergoing Aspherical Explosions}

These stars undergo Fe-core collapse and become Type II-P (plateau)
supernovae (SNe II-P) if the red-supergiant-size H-rich envelope remains,
and Type Ibc supernovae if the H-rich envelope has been stripped off
by a stellar wind or Roche-lobe overflow.  These SNe are the major
sources of heavy elements from C to the Fe-peak.

Their yields depend on the progenitor's mass $M$, metallicity, and the
explosion energy $E$.  From the comparison between the observed and
calculated spectra and light curves of supernovae, we can estimate $M$
and $E$ as shown in Figure~\ref{figME} (\cite{nomoto2006}).

Three SNe (SNe~1998bw, 2003dh, and 2003lw) are associated with long
Gamma-Ray Bursts (GRBs) (e.g., \cite{woo06}).  The progenitors of
these GRB-SNe tend to be more massive than $\sim 30 M_\odot$.  Also
GRB-SNe are all very energetic with the kinetic energy $E$ exceeding
$10^{52}$\,erg, more than 10 times the kinetic energy of normal
core-collapse SNe.  Here we use the term 'Hypernova (HN)' to describe
such hyper-energetic supernovae with $E_{51} = E/10^{51}$ erg $\gsim
10$ (Fig.\ref{figME}; Nomoto \etal\ 2004, 2006; \cite{koba06}).

\subsection{Supernova and Hypernova Yields}

\begin{figure}[t]
\begin{center}
\includegraphics[width=10.0cm]{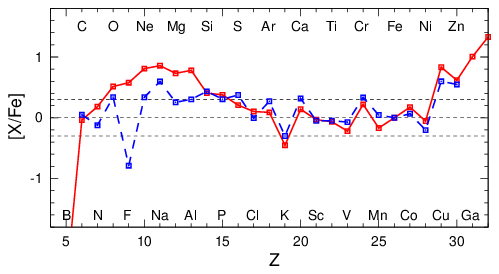}
\includegraphics[width=10.0cm]{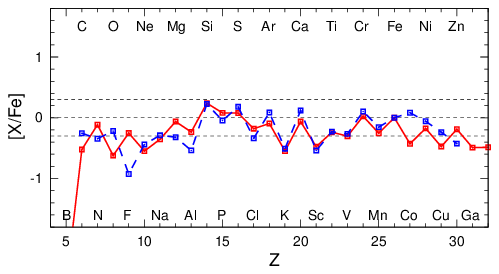}
\end{center}
\caption{
({\sl Upper}): Comparison of yields for $M=15 M_\odot$, $Z =$ 0.02,
and $E = 1 \times 10^{51}$ erg (red-solid: \cite{nomoto2006};
blue-dashed: \cite{limo2000}).
({\sl Lower}): Same as upper, but for $M=25 M_\odot$.
}
\label{fz02}
\end{figure}

Theoretical models of stellar evolution depend on the treatment of
complicated physical processes, such as mixing due to convection and
rotation, convective overshooting, mass loss, etc.  Thus SN yields
obtained by various groups are not necessarily in agreement.  Figure
\ref{fz02} compares the yields of the models with $E = 1 \times
10^{51}$ erg and $Z =$ 0.02 for $M=15 M_\odot$ (upper) and $M=25
M_\odot$ (lower) (\cite{nomoto2006}; \cite{limo2000}). These models
include mass loss but not rotation.  It is seen that two yields are in
good agreement.

\begin{figure}[t]
\begin{center}
\includegraphics[width=10.0cm]{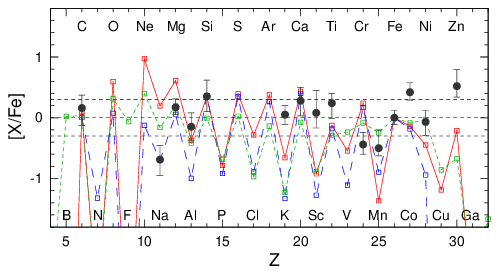}
\includegraphics[width=10.0cm]{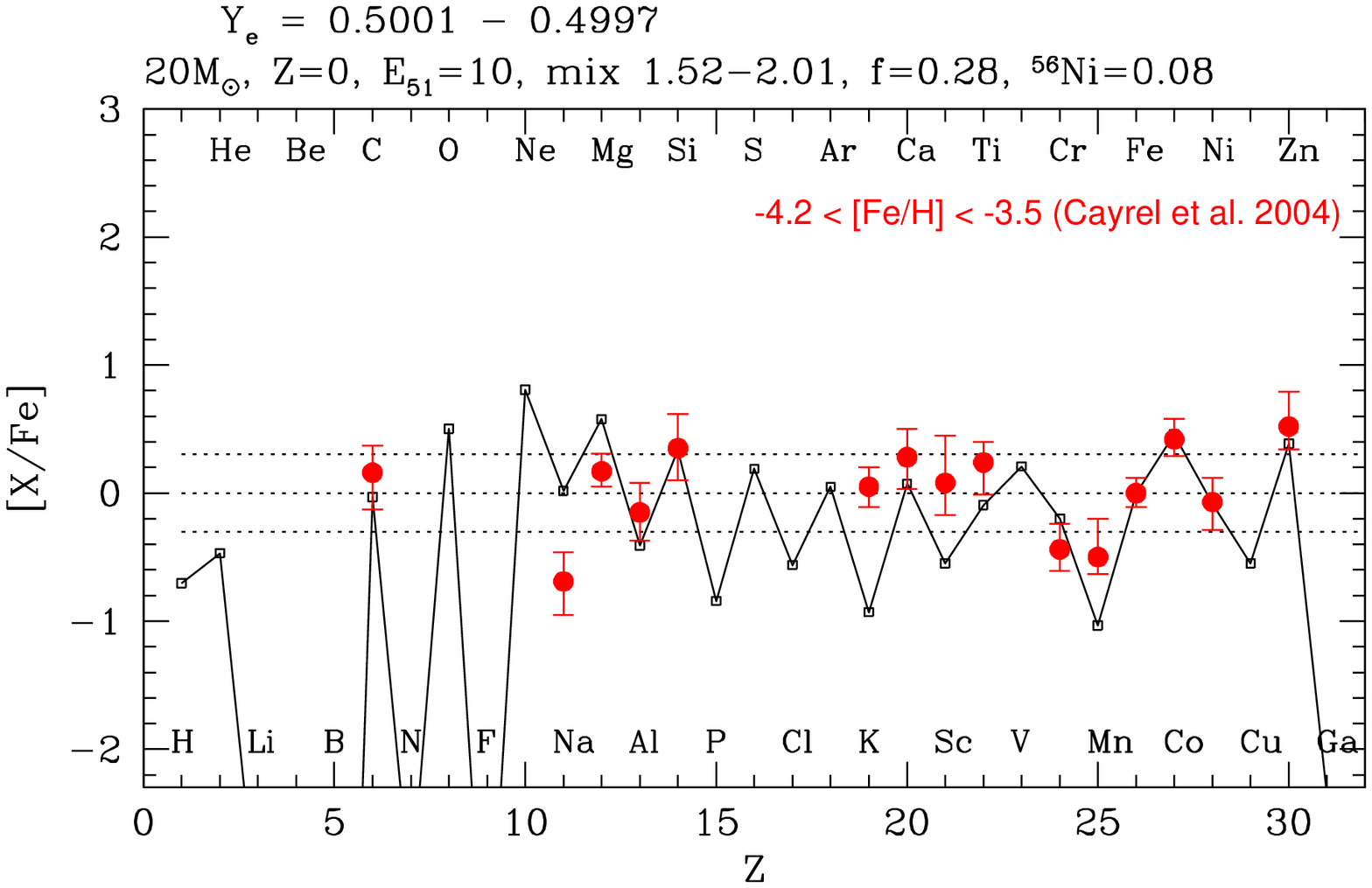}
\end{center}
\caption{
Averaged elemental abundances of stars with [Fe/H] $= -3.7$
(\cite{cayrel2004}) compared with yields for $M=20 M_\odot$ and $Z=$
0.0 ({\sl upper}: \cite{nomoto2006}; \cite{limo2000};
\cite{heger2008}) and the hypernova yield ({\sl lower}: 20 $M_\odot$,
$E_{51} =$ 10).
}
\label{fz00_comp}
\end{figure}

Figure \ref{fz00_comp} (upper) compares the yields of models with
$E = 1 \times 10^{51}$ erg and $Z =$ 0.00 for $M=20 M_\odot$ between
the three groups (\cite{nomoto2006}; \cite{limo2000};
\cite{heger2008}).  These three yields are in good agreement.  The
smooth pattern in Heger \& Woosley (2008) is due to the
mixing-fallback effect (\cite{ume02}) being taken into account in their model.  
This implies that the difference in the treatment of such mixing-fallback
during the explosion is larger than other differences in presupernova
models among the three groups.

Figure \ref{fz00_comp} (upper) also compares the observed averaged
abundance pattern of the EMP stars (\cite{cayrel2004}) with the three
theoretical models.  We note that theoretical predictions of Zn, Co,
Ti/Fe are much smaller than the observed ratios.  The underproduction
of these elements relative to Fe is much improved in the hypernova
models (lower); this suggests that hypernovae play an important role in
the chemical enrichment during early galactic evolution.

In the following section, therefore, we focus on nucleosynthesis in
the high energy jet-induced explosions (\cite{tomi07a};
\cite{tomi08}).  In the jet-like explosion, fallback can occur even
for $E > 10^{52}$ erg.

\subsection{Nucleosynthesis in Jet-Induced Explosions and GRB-SN Connection}

The observed late-time spectra indicate that the explosions of these
Fe-core collapse supernovae are quite aspherical (\cite{mae08};
\cite{moj08}).  The extreme case is the hyper-aspherical explosions
induced by relativistic jet(s) as seen in the GRB-SNe.

\begin{figure}[t]
\begin{center}
\includegraphics[width=8.5cm]{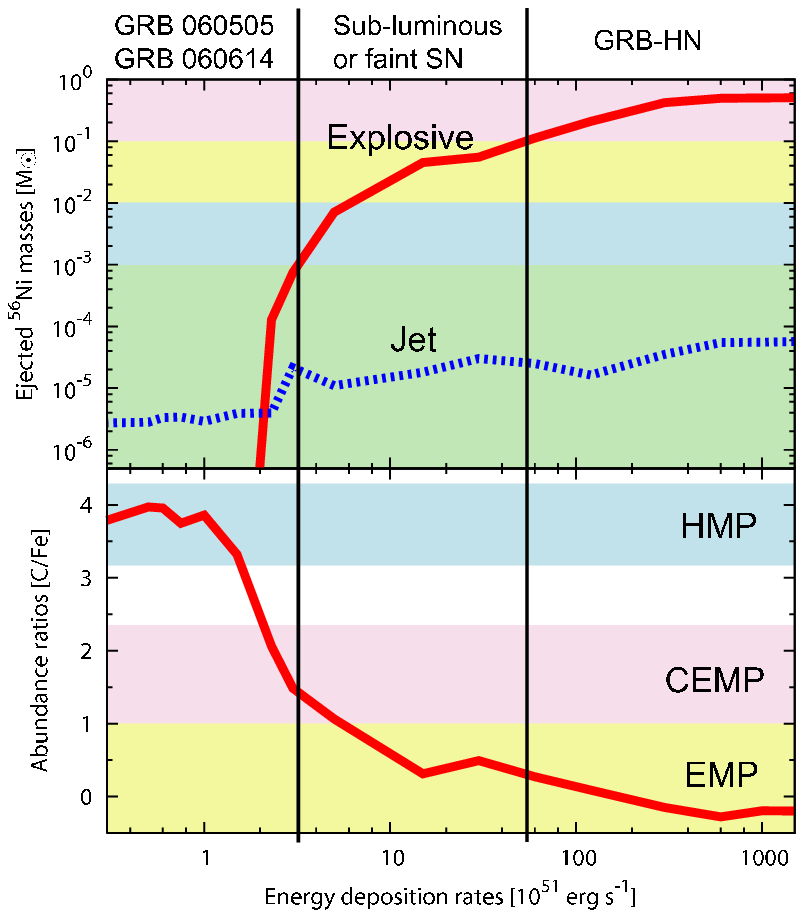}
\caption{{\it Upper}: The ejected \Nifs\ mass ({\it red}: explosive
 nucleosynthesis products, {\it blue}: the jet contribution) as a
 function of the energy deposition rate (\cite{tomi07a}). The
 background color shows the corresponding SNe ({\it red}: GRB-HNe,
 {\it yellow}: sub-luminous SNe, {\it blue}: faint SNe, {\it green}:
 GRBs~060505 and 060614).  Vertical lines divide the resulting SNe
 according to their brightness.  {\it Lower}: the dependence of
 abundance ratio [C/Fe] on the energy deposition rate. The background
 color shows the corresponding metal-poor stars ({\it yellow}: EMP,
 {\it red}: CEMP, {\it blue}: HMP stars).}
\label{fig:EdotNi}
\end{center}
\end{figure}

Recent studies of nucleosynthesis in jet-induced explosions have
revealed the connection between GRBs and EMP stars as summarized in
Figure~\ref{fig:EdotNi} (\cite{tomi07a}; \cite{tomi08}).  In this
model for the $40\Msun$ star, the jets are injected at a radius $R_0
\sim 900$ km with energy deposition rates in the range 
$\Edep\equiv\Ed/10^{51}{\rm ergs\,s^{-1}}=0.3-1500$.  The diversity of
$\Ed$ is consistent with the wide range of the observed isotropic
equivalent $\gamma$-ray energies and timescales of GRBs (\cite{ama06}
and references therein).  Variations of activities of the central
engines, possibly corresponding to different rotational velocities or
magnetic fields, may well produce the variation of $\Ed$.

The ejected $\Mni$ depends on $\Ed$ as follows
(Fig.~\ref{fig:EdotNi}).  Generally, higher $\Ed$ leads to the
synthesis of larger $\Mni$ in explosive nucleosynthesis because of
higher post-shock densities and temperatures (e.g., \cite{mae03};
\cite{nag06}).  If $\Edep\gsim60$, we obtain $\Mni$ $\gsim0.1\Msun$,
which is consistent with the brightness of GRB-HNe.  Some C+O core
material is ejected along the jet direction, but a large amount of
material along the equatorial plane falls back.

For $\Edep\gsim60$, the remnant mass is initially $M_{\rm rem}^{\rm
start}\sim1.5\Msun$ and grows as materials is accreted from the
equatorial plane.  The final BH mass is generally larger for smaller
$\Ed$.  The final BH masses range from $M_{\rm BH}=10.8\Msun$ for
$\Edep=60$ to $M_{\rm BH}=5.5\Msun$ for $\Edep=1500$, which are
consistent with the observed masses of stellar-mass BHs
(\cite{bai98}).  The model with $\Edep=300$ synthesizes
$\Mni\sim0.4\Msun$, and the final mass of BH left after the explosion
is $M_{\rm BH}=6.4\Msun$.

For low energy deposition rates ($\Edep<3$), in contrast, the ejected
\Nifs\ masses ($\Mni<10^{-3}\Msun$) are smaller than the upper limits
for GRBs~060505 and 060614 (\cite{del06}; \cite{fyn06}; \cite{gal06}).

\begin{figure}[t]
\begin{center}
\includegraphics[width=10cm]{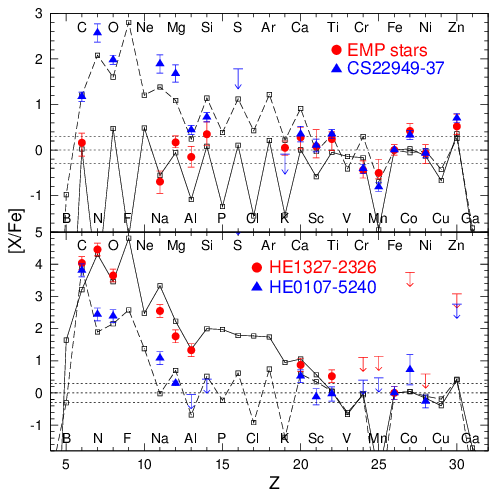}
\caption{
 A comparison of the abundance patterns between metal-poor 
 stars and models (\cite{tomi07a}).
 {\it Upper}: typical EMP stars ({\it red dots}, \cite{cayrel2004}) and 
 CEMP ({\it blue triangles}, CS~22949--37, \cite{dep02}) and models
 with $\Edep=120$ ({\it solid line}) and $=3.0$ ({\it dashed line}).
 {\it Lower}: HMP stars:
 HE~1327--2326, ({\it red dots}, e.g., \cite{fre05}), 
 and HE~0107--5240, ({\it blue triangles}, \cite{chr02, bes05}) and models 
 with $\Edep=1.5$ ({\it solid line}) and $=0.5$ ({\it dashed line}).
\label{fig:HMP}}
\end{center}
\end{figure}

If the explosion is viewed from the jet direction, we would observe
the GRB without SN re-brightening. This may be the situation for
GRBs~060505 and 060614.  In particular, for $\Edep<1.5$, \Nifs\ cannot
be synthesized explosively, and the jet component of the Fe-peak
elements dominates the total yields (Fig.~\ref{fig:HMP}). The
models eject very little $\Mni$ ($\sim10^{-6}\Msun$).

For intermediate energy deposition rates ($3\lsim\Edep<60$), the
explosions eject $10^{-3}\Msun \lsim \Mni <0.1\Msun$, and the final BH
masses are $10.8\Msun\lsim M_{\rm BH}< 15.1\Msun$. The resulting SN is
faint ($\Mni <0.01\Msun$) or sub-luminous ($0.01\Msun \lsim \Mni
<0.1\Msun$).

\subsection{Abundance Patterns of Extremely Metal-Poor Stars}

The abundance ratio [C/Fe] depends on $\Ed$ as follows.  Lower $\Ed$
yields larger $M_{\rm BH}$ and thus larger [C/Fe], because the infall
reduces the amount of inner core material (Fe) relative to that of
outer material (C) (see also \cite{mae03}).  As in the case of $\Mni$,
[C/Fe] changes dramatically at $\Edep\sim3$.

The observed abundance patterns of EMP stars are good indicators of SN
nucleosynthesis because the Galaxy was effectively unmixed at [Fe/H]
$< -3$ (e.g., \cite{tum06}). They are classified into three groups
according to [C/Fe]:

(1) [C/Fe] $\sim 0$, normal EMP stars ($-4<$ [Fe/H] $<-3$, e.g., 
    \cite{cayrel2004});

(2) [C/Fe] $\gsim+1$, Carbon-enhanced EMP (CEMP) stars ($-4<$ [Fe/H]
    $<-3$, e.g., CS~22949--37, \cite{dep02});

(3) [C/Fe] $\sim +4$, hyper metal-poor (HMP) stars ([Fe/H] $<-5$,
    e.g., HE~0107--5240, \cite{chr02}; \cite{bes05}; HE~1327--2326,
    \cite{fre05}).

Figure \ref{fig:HMP} shows that the abundance patterns of the averaged
normal EMP stars, the CEMP star CS~22949--37, and the two HMP stars
(HE~0107--5240 and HE~1327--2326) are well reproduced by the models
with $\Edep=120$, 3.0, 1.5, and 0.5, respectively. The model for the
normal EMP stars ejects $\Mni\sim0.2\Msun$, i.e., a factor of 2 less
than SN~1998bw. On the other hand, the models for the CEMP and the HMP
stars eject $\Mni\sim8\times10^{-4}\Msun$ and $4\times 10^{-6}\Msun$,
respectively, which are always smaller than the upper limits for
GRBs~060505 and 060614.  The N/C ratio in the models for CS~22949--37
and HE~1327--2326 is enhanced by partial mixing between the He and H
layers during presupernova evolution (\cite{iwamoto2005}).

To summarize, (1) the explosions with large energy deposition rate,
$\Ed$, are observed as GRB-HNe, and their yields can explain the
abundances of normal EMP stars, and (2) the explosions with small
$\Ed$ are observed as GRBs without bright SNe and can be responsible
for the formation of the CEMP and the HMP stars.  We thus propose that
GRB-HNe and GRBs without bright SNe belong to a continuous series of
BH-forming massive stellar deaths with relativistic jets of
different $\Ed$.

\section{$90 - 140 M_\odot$ Stars undergoing Pulsational Nuclear Instabilities}

These massive stars undergo nuclear instabilities and associated
pulsations ($\epsilon$-mecha\-nism) at various nuclear burning stages.
Because of the large contribution of radiation pressure in these
stars, dynamical stability is very close to neutral.  Even a
slight contribution of electron-positron pair creation affects the
stability.  Thus the pulsation behavior is sensitive to their
mass, mass loss rate, and metallicity.

To determine the above upper mass limit, the non-adiabatic stability 
of massive Pop III ($Z = 0$) stars has been analyzed
(\cite{ibrahim1981}; \cite{baraffe2001}; \cite{nom03}).  As summarized
in Table~\ref{tab:pop3} (\cite{nom03}), the critical mass of a Pop III
star is $128M_\odot$, while that of a Pop I star is $94M_\odot$.  This
difference stems from the very compact structure (with high central
temperature) of Pop III stars.  Stars more massive than the critical
mass will undergo pulsation and mass loss. We note that the
$e$-folding time of instability is much longer for Pop III stars than
Pop I stars with the same mass, and thus the mass loss rate is much
lower.  Thus, massive Pop III stars could survive the instabilities
without losing much mass.

\begin{table}[t]

\caption{Stability of Pop III and Pop I massive stars: $\bigcirc$
and $\times$ indicate that the star is stable or unstable,
respectively.  The $e$-folding time for the fundamental mode is shown
after $\times$ in units of $10^4$yr (\cite{nom03}).}
\medskip
\begin{center}
\footnotesize
\begin{tabular}{ccccccc}
\hline
 $M(M_\odot)$ & 80 & 100 &120 & 150 & 180 & 300 \\ \hline
 Pop III & $\bigcirc$ & $\bigcirc$ & $\bigcirc$ & $\times$ (9.03) & 
 $\times$ (4.83) & $\times$ (2.15) \\ 
 Pop I & $\bigcirc$ & $\times$ (7.02) & $\times$ (2.35) & 
 $\times$ (1.43) & $\times$ (1.21) & $\times$ (1.71) \\ \hline
\end{tabular}
\end{center}
\label{tab:pop3}
\end{table}

Massive Pop III stars are formed through mass accretion, starting from a
tiny core through collapse (e.g., \cite{yoshida08}).  Such an
evolution with mass accretion starting from $M \sim 1 M_\odot$ has
recently been studied by Ohkubo \etal\ (2006, 2008).

\begin{figure}[t]
\begin{center}
\includegraphics[width=10cm]{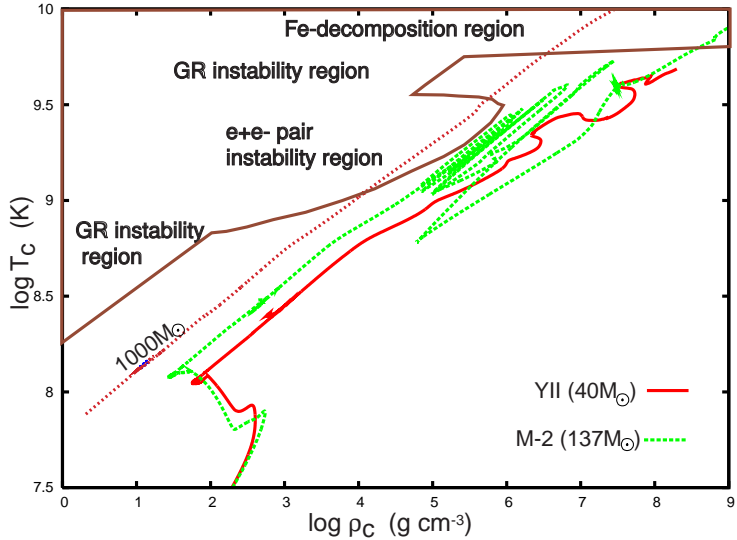}
\end{center}
\caption{
Evolutionary tracks of the central temperature and central density of
very massive stars (\cite{Ohku09}).  The numbers in brackets are
the final masses for models YII and M-2.  The 1000$M_{\odot}$ stars 
(\cite{Ohku06}) are also shown.
}
\label{fig:vms_rhot}
\end{figure}

Figure~\ref{fig:vms_rhot} shows the evolutionary tracks of the central
density and temperature in the later phases.  For the models with mass
accretion (M-2, YII), the central entropy in the early stage is low,
corresponding to the small stellar mass.  During the main-sequence
phase, the stellar mass increases to $\sim 40M_{\odot}$ (YII) and
$\sim 137 M_{\odot}$ (M-2), and the density-temperature track shifts
toward corresponding higher entropy through hydrogen burning.  

The star YII ($\sim 40M_{\odot}$) ends its life in Fe-core collapse to
form a black hole.  The star M-2, whose final mass is 137$M_{\odot}$,
undergoes nuclear instability due to oxygen and silicon burning and
pulsates, as seen in Figure~\ref{fig:vms_rhot}.  Such pulsations
are seen in stars with masses $90M_{\odot} \lsim M \lsim 140M_{\odot}$
(\cite{nom05}; \cite{Woos07}; \cite{ume08}; \cite{Ohku09}).  In the
extreme case, the pulsation could induce dynamical mass ejection and
optical brightening as might be observed in the brightest SN 2006gy
(\cite{Woos07}).  The nuclear energy released is not sufficient to explode
the whole star.  After several oscillations, the star finally
collapses to form a black hole.

\begin{figure}[t]
\begin{center}
\includegraphics[width=10cm]{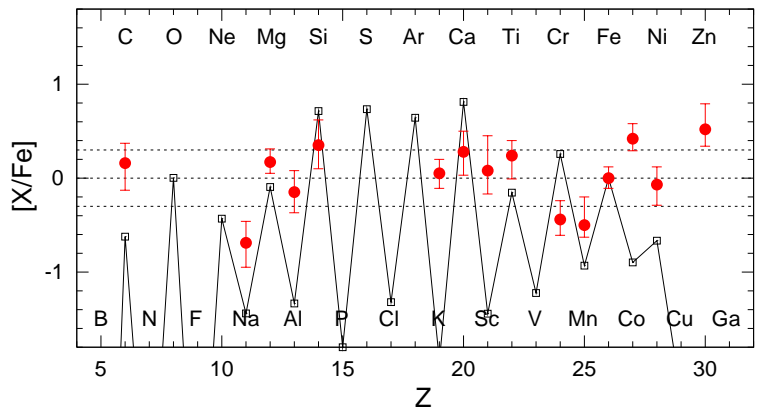}
\end{center}
\caption{
Comparison of the abundance patterns in the pair-instability supernova 
model ($M = 200 M_\odot$; \cite{ume02}) and extremely metal-poor
stars (\cite{cayrel2004}).
\label{fig:pisn_emp}}
\end{figure}

\section{140 - 300 $M_\odot$ Stars undergoing Pair-Instability Supernovae}

These very massive stars (VMS) undergo pair-creation instability and
are disrupted completely by explosive oxygen burning, as 
pair-instability supernovae (PISNe) (e.g., \cite{barkat1967};
\cite{arnett1996}; \cite{ume02}; \cite{heger2002}).

The abundance patterns of the ejected material for the 200 $M_\odot$
star (\cite{ume02}) are compared with EMP stars (\cite{cayrel2004}) in
Figure~\ref{fig:pisn_emp}.  It is clear that PISN ejecta cannot be
consistent with the large C/Fe observed in HMP stars and other C-rich
EMP stars.  Also, the abundance ratios of iron-peak elements ([Zn/Fe] $<
-0.8$ and [Co/Fe] $< -0.2$) in the PISN ejecta cannot explain the
large Zn/Fe and Co/Fe ratios in typical EMP stars.

Therefore, the supernova progenitors that are responsible for the
formation of EMP stars are most likely in the range of $M \sim 20 -
140$ $M_\odot$, but not more massive than 140 $M_\odot$.  The absence
of any indication of PISNe in EMP stars might imply that 140 - 300
$M_\odot$ stars might not have formed from accretion in Pop III stars,
or underwent significant mass loss, thus evolving into Fe
core-collapse.

\section{Very Massive Stars with $M > 300 M_\odot$ and Intermediate Mass Black Holes}

It is possible that the First Stars were even more massive than $\sim
300 M_\odot$, if rapid mass accretion continues during the whole
main-sequence phase of Pop III stars (Ohkubo \etal\ 2006, 2008).
[Another possible scenario for any metallicity is that VMSs are formed
by merging of less massive stars in the environment of very dense star
clusters (e.g., \cite{ebisu01}; \cite{zwart07})].

Such massive stars undergo core-collapse (CVMS: core-collapse VMS) as
seen from the 1000 $M_\odot$ star track in Figure~\ref{fig:vms_rhot}.
If such stars formed rapidly rotating black holes, jet-like mass
ejection could produce processed material (\cite{Ohku06}).  In
fact, for moderately aspherical explosions, the patterns of
nucleosynthesis match the observational data of both intracluster
medium and M82 (\cite{Ohku06}).  This result suggests that explosions
of CVMS contribute significantly to the chemical evolution of gases in
clusters of galaxies.  For Galactic halo stars, predicted [O/Fe]
ratios are smaller than the observational abundances.  This result may
support the view that Pop III CVMS could be responsible for the origin
of intermediate mass black holes (IMBH).

\bigskip

This research has been supported in part by World Premier
International Research Center Initiative (WPI Initiative), MEXT,
Japan, and by the Grant-in-Aid for Scientific Research of the JSPS
(18104003, 18540231, 20540226) and MEXT (19047004, 20040004).

\end{document}